\documentclass[preprint2]{aastex}
\usepackage{emulateapj5}
\usepackage{onecolfloat5}
\shorttitle{Bow Shocks from Neutron Stars}
\slugcomment{Submitted to the Astrophysical Journal, 30 Dec 2001; 
	     revised 20 Mar 2002.}

\def\Eref#1{Eqn.~\ref{Eqn:#1}}
\def\Sref#1{\S\ref{Sec:#1}}
\def\Fref#1{Fig.~\ref{Fig:#1}}
\def\Tref#1{Table~\ref{Table:#1}}
\newcommand{\persec}{s$^{-1}$}
\newcommand{\Halpha}{H$\alpha$}
\newcommand{\nHI}{n_{\rm H\,\scriptscriptstyle{I}}}
\newcommand{\gHa}{g_{{\rm H}\alpha}}

\newcommand{\nH}{n_{\rm H}}
\newcommand{\mH}{m_{\rm H}}
\newcommand{\nA}{n_{\rm A}}
\newcommand{\gammaH}{\gamma_{\rm H}}
\newcommand{\GN}{Guitar Nebula}
\newcommand{\Edot}{$\dot{E}$}
\newcommand{\mas}{milliarcsecond}

\begin{document}
\twocolumn[
\title{Bow Shocks from Neutron Stars: Scaling Laws and \\ 
	HST Observations of the Guitar Nebula}
\author{S. Chatterjee \& J. M. Cordes}
\affil{Department of Astronomy, Cornell University, Ithaca, NY 14853}
\email{shami@astro.cornell.edu, cordes@astro.cornell.edu}

\begin{abstract}

The interaction of high-velocity neutron stars with the interstellar
medium produces bow shock nebulae, where the relativistic neutron star
wind is confined by ram pressure.  We present multi-wavelength
observations of the Guitar Nebula, including narrow-band \Halpha\
imaging with HST/WFPC2, which resolves the head of the bow shock. The
HST observations are used to fit for the inclination of the pulsar
velocity vector to the line of sight, and to determine the combination
of spindown energy loss, velocity, and ambient density that sets the
scale of the bow shock.  We find that the velocity vector is most
likely in the plane of the sky.  We use the \GN\ and other observed
neutron star bow shocks to test scaling laws for their size and
\Halpha\ emission, discuss their prevalence, and present criteria
for their detectability in targeted searches.  The set of \Halpha\ bow
shocks shows remarkable consistency, in spite of the expected
variation in ambient densities and orientations.  Together, they
support the assumption that a pulsar's spindown energy losses are
carried away by a relativistic wind that is indistinguishable from
being isotropic.  Comparison of \Halpha\ bow shocks with X-ray and
nonthermal, radio-synchrotron bow shocks produced by neutron stars
indicates that the overall shape and scaling is consistent with the
same physics.  It also appears that nonthermal radio emission and
\Halpha\ emission are mutually exclusive in the known objects and
perhaps in all objects.

\end{abstract}

]

\section{INTRODUCTION}\label{Sec:intro}

Bow shocks are observed on a wide variety of astrophysical scales,
ranging from planetary magnetospheres \citep{bowPlanet} and OB-runaway
stars \citep{bowOB} to merging galaxy clusters \citep{bowGC}.  They
have been invoked in models for beamed gamma-ray bursts
\citep{bowGRB}, protostellar outflows \citep{OLSM01}, ultra-compact
\ion{H}{2} regions \citep{bowHII} and the interaction of the solar
wind with the interstellar medium \citep{BKK71}.  Some of the most
spectacular bow shock nebulae are those associated with neutron stars.
The spindown of neutron stars (NS) transfers rotational kinetic energy
into the interstellar medium, and bow shock nebulae provide a way to
probe these energetic environments, revealing the properties of both
the NS relativistic winds and the ambient interstellar medium (ISM).

The Guitar Nebula is one such visually striking bow shock nebula,
produced by the radio pulsar B2224+65. The nebula was discovered in
\Halpha\ observations using the 5-m Hale Telescope at Palomar
Observatory \citep{CRL93}.  Until recently, there were only two other
NS with bow shocks observed in \Halpha: B1957+20 \citep{KH88} and
B0437$-$4715 \citep{BBM+95}.  However, \Halpha\ bow shocks have
recently been discovered around a radio quiet NS, RX~J1856.5$-$3754
\citep{VK01}, as well as another ordinary radio pulsar, B0740$-$28
\citep{JSG02}.  More such nebulae will no doubt be discovered as
targeted searches become more sensitive and cover more objects.
Additionally, there are some bow shock nebulae which are detected at
radio/X-ray wavelengths, but not in \Halpha, associated with known
radio pulsars as well as radio-quiet NS: for example, the ``Duck''
\citep{FK91} is associated with the pulsar B1757$-$24 and the
supernova remnant (SNR) G5.4$-$1.2, while the cometary nebula in the
SNR IC~443 is produced by a radio quiet NS \citep{OCW+01}.


In this paper, we derive scaling relationships for bow shock
parameters based on the underlying physical processes, and discuss the
requirements for such nebulae to be detectable.  A detailed analysis
of the Guitar Nebula is presented based on multi-wavelength
observations, including high resolution Hubble Space Telescope (HST)
observations which resolve the fine structure of the bow shock.
Through model fitting, constraints are obtained on the density of the
ambient ISM and the inclination of the NS velocity vector to the line
of sight (LOS).  The published data on NS bow shocks are consolidated
with the parameters for the Guitar Nebula in order to test the derived
scaling relationships for \Halpha\ bow shocks.  We verify that the
stand-off angle for plerionic bow shock nebulae detected at radio and
X-ray wavelengths also scales in the same manner.  In testing these
scaling laws, it is apparent that the current state of available
information is inadequate, especially for radio-detected nebulae,
leading to constraints on the ISM that are not very strong.  We
identify the sources of uncertainty, and suggest future observations
that will be able to refine current constraints as well as confirm or
refute physical scenarios relating to the formation of \Halpha\ and
plerionic bow shock nebulae.

The paper is organized as follows: the expected scaling relationships
for bow shock parameters are derived in \Sref{scale}, along with an
analysis of the detectability criteria for bow shock nebulae.
Multi-wavelength observations of the \GN\ are described in \Sref{GN}.
Since the bright head of the \GN\ is not resolved by ground-based
optical images, high resolution HST observations are required to
obtain bow shock parameters; these observations, and the associated
model fitting, are described in \Sref{HST}. In \Sref{test}, the
derived scaling laws are tested for known NS bow shocks, with
consolidated parameters for both \Halpha\ and radio/X-ray bow shocks.
Finally, in \Sref{end}, we summarize our results, discuss their
implications and identify future lines of inquiry.

\section{SCALING LAWS FOR BOW SHOCK NEBULAE}\label{Sec:scale}

Bow shock nebulae are produced when NS move supersonically through the
ambient ISM, creating a shocked layer where ram pressure balance is
established between the relativistic NS wind and the medium.  As such,
bow shock characteristics such as the stand-off angle between the NS
and the tip of the bow shock should scale with the spindown energy
loss rate of the NS and the density of the medium. Pulsars with
detected \Halpha\ bow shocks (J0437$-$4715, B0740$-$28, B1957+20 and
B2224+65) have a wide range of properties, with \Edot\ ranging over
two orders of magnitude, and velocities and distances spanning an
order of magnitude, as detailed in \Sref{test}.  This leads to the
question of how the bow shock scales with changes in the properties of
the NS and the ambient medium.

\subsection{Scaling of the Stand-Off Angle}

\citet{W96} provides an exact analytic solution for a stellar wind bow
shock in the thin-shell limit, a problem first solved numerically by
\citet{BKK71}.  In this analytic treatment, ram pressure balance
between the stellar wind and the ambient medium determines the
stand-off radius of the bow shock ($R_0$).  For a star moving with
velocity $v_{\ast}$ through a uniform medium of density $\rho_a$, with
a stellar wind mass loss rate $\dot{m}_w$ and wind velocity $v_w$, the
condition for ram-pressure balance is
\begin{equation}
\rho_a v^2_{\ast} = \frac{\dot{m}_w v_w}{4 \pi R_0^2},
\end{equation}
leading to an expression for the stand-off radius,
\begin{equation} 
R_0 = \left(\frac{\dot{m}_w v_w}{4 \pi \rho_a v^2_{\ast}}\right)^{1/2}.
\end{equation} 

For a relativistic NS wind, the momentum outflow per unit area
${\dot{m}_w v_w}$ can be recast as $\dot{E}/c$ under the assumption
that the spindown energy loss is carried away by the relativistic
wind. Then the stand-off radius becomes
\begin{equation}
R_0 = \left(\frac{\dot{E}}{4 \pi c \rho_a v^2_{\ast}}\right)^{1/2}. 
\label{Eqn:R0} 
\end{equation}

The substitution of a relativistic NS wind in place of an isotropic
stellar wind may be problematic, since the NS wind is at best only
quasi-isotropic.  Some authors have cited the possibility of an
alignment of the NS spin axis and proper motion \citep[e.g.][]{SP98},
and two pulsars (the Crab and Vela) apparently conform to this picture
\citep{LCC01}, although it might not be a general phenomenon
\citep{DRR99}.  In case of alignment, the rotation-averaged
relativistic wind, while not isotropic, would be axisymmetric about
the velocity vector, a situation which adds only one extra scaling
parameter to the bow shock description \citep{W00}.  If there is no
such alignment, so that the rotation vector is skewed with respect to
both the velocity and the magnetic field, the behavior of the
relativistic wind depends on the relative orientations of these
vectors and the opening angle of the wind: for a range of
orientations, rotational averaging leads to a quasi-isotropic
situation relative to the velocity vector.  Specifically, even for
B2224+65, a relatively slow-spinning pulsar with a small stand-off
radius $R_0$, the light cylinder radius (at which point the
rotation-averaged NS wind has spread out over a large solid angle) is
$\sim$ a few $\times 10^{-6} R_0$.  In the absence of additional
information, we adopt an isotropic description for the relativistic
wind.

The observed stand-off angle $\theta_0$ is simply the projected
stand-off radius at the NS distance:
\begin{equation}
\theta_0 = R_0 \sin i / D,  \label{Eqn:deftheta}
\end{equation}
where $i$ is the inclination angle to the LOS, so that $i=\pi/2$
corresponds to a velocity vector in the plane of the sky, and
$i=0$ corresponds to motion towards the observer.  The measurable
transverse velocity of the NS, $v_\perp$, is related to the actual
velocity $v_\ast$ by the same projection factor:
\begin{equation}
v_\perp = \mu D = v_\ast \sin i,
\end{equation} 
where $\mu$ is the measured angular proper motion.
The ambient density of the ISM, $\rho_a$, is the product of the number
density of the medium ($\nH$), the mass of the H atom ($\mH$) and the
equivalent molecular weight of the hydrogen and helium mixture
($\gammaH$ = 1.37 for cosmic abundances, with 27\% He by mass).  
Defining $\nA$, the density in atomic mass units per unit volume,
\begin{equation}
\rho_a = \nH \gammaH \mH = \nA \mH.
\end{equation}
Making the appropriate substitutions in \Eref{R0}, and
factoring in terms of constants, unknown quantities and observables,
we obtain: 
\begin{equation}
\theta_0 = \left(4 \pi \mH c \right)^{-1/2}
	     \left(\frac{\sin^2 i}{\nA^{1/2}}\right)
	     \left(\frac{\dot{E}^{1/2}}{\mu D^2}\right);
\label{Eqn:theta0}
\end{equation}
the last factor comprises the quantities \Edot, $D$ and $\mu$, which
are (at least in principle) directly measurable.  \Eref{theta0} can be
evaluated numerically, for $\theta_0$ in \mas, $\dot{E}_{33}$ in
$10^{33}$ erg~\persec, $D$ in kpc, $\mu_{100}$ in 100 \mas\ yr$^{-1}$
and $\nA$ in cm$^{-3}$:
\begin{equation}
\theta_0 = 56.3 \,{\rm mas}
	     \left(\frac{\sin^2 i}{\nA^{1/2}}\right)
	     \left(\frac{\dot{E}_{33}^{1/2}}{\mu_{100} D_{\rm kpc}^2}\right).
\label{Eqn:numtheta}
\end{equation} 

\subsection{Scaling of the \Halpha\ Flux}

Bow shocks emit \Halpha\ photons when the partially neutral ISM
encounters the shock front, traveling at the NS velocity $v_\ast$.
The neutral atoms are not immediately affected by the shock
transition, but undergo collisional excitation and ionization in the
hot post-shock flow due to encounters with shocked electrons and
protons \citep{CR78}.  \Halpha\ emission is proportional to the
probability of an excitation before ionization, which is given by the
ratio of the excitation rate $q_{\rm ex}$ to the ionization rate
$q_{\rm i}$ \citep{R91}. The observed rate of \Halpha\ emission
($F_\alpha$) from a nebula which subtends a solid angle $\Delta
\Omega$ is given by:
\begin{equation}
F_\alpha = \frac{q_{\rm ex}}{q_{\rm i}}\; \nHI v_{\ast}\; 
\frac{\Delta \Omega}{4 \pi \gHa},  \label{Eqn:falpha}
\end{equation}
where $\gHa$ is the extinction correction for \Halpha\ ($\approx
3$ for the Guitar Nebula), and \citet{R91} gives $q_{\rm ex}/q_{\rm i}
\approx 0.2$ \Halpha\ photons per neutral atom.

The solid angle subtended by the bow shock nebula $\Delta \Omega \sim
(\eta R_0/D)^2$, where $\eta$ is a geometric factor which depends on
the NS velocity. \citet{CRL93} use $\eta \propto v_\ast$, which is
physically plausible, since a faster-moving NS sweeps up a larger
volume of ISM within the shock front during the ionization timescale,
so that $\Delta \Omega \propto v_\ast^2$.  To retain generality, we
use $\eta \propto v_{\ast}^{\beta}$, which implies $\Delta \Omega
\propto v_{\ast}^{2\beta} (R_0/D)^2$. In addition, defining the neutral
fraction $X = \nHI/\nH$ leads to $\nHI = X\nH \propto X \rho_a$, and
substituting for $\Delta \Omega$ and $\nHI$ in \Eref{falpha}, 
\begin{equation}
F_\alpha \propto X \nH\, v_\ast (\eta R_0/D)^2.
\end{equation}
The \Halpha\ photons are emitted isotropically by collisionally
excited neutral atoms, so that geometric projection of the stand-off
radius and the nebula itself does not affect the total observed flux
$F_\alpha$, but projection effects have to be accounted for to express
the actual NS velocity $v_\ast$ in terms of the observables $\mu$ and
$D$. Thus the \Halpha\ flux can be expressed in terms of observable
quantities ($\theta_0, \mu, D$) and unknowns ($\sin i, \nH, X$):
\begin{equation}
F_\alpha \propto X \nH\, \theta_0^2 {v_\ast}^{2\beta+1} \;
	 \propto X \nH\, \theta_0^2
	 \left(\frac{\mu D}{\sin i}\right)^{2\beta+1}.
\label{Eqn:f1}
\end{equation}

Alternatively, using the definition of $R_0$ from \Eref{R0}, a
scaling relationship can be derived for the observed \Halpha\ flux in
terms of measurable quantities and the unknown neutral fraction:
\begin{equation}
F_\alpha \propto \frac{X \dot{E} v_{\ast}^{2\beta-1}}{D^2},
\label{Eqn:f2}
\end{equation}
which reduces to the expression derived by \citet{CRL93} for $\beta =
1$.

\subsection{Detectability of \Halpha\ Bow Shock Nebulae}

For bow shock nebulae, the observed \Halpha\ emission comes from the
outer shocked layer, where the neutral medium is swept up and
collisionally excited. There is also an inner shocked region which
encloses the relativistic NS wind and may produce synchrotron
radiation detectable in X-ray or radio observations.  \Halpha\
emission requires the existence of a significant neutral component in
the medium, which is true for only $\sim 20\%$ of the ISM
\citep{MO77}.  Thus the non-detection of a nebula in \Halpha\ may
reflect pre-ionization of the ISM by supernovae, or thermal radiation
from the NS or shocked gas, and places an upper limit on the neutral
fraction present in the ambient medium.  This is especially applicable
for nebulae detected at radio/X-ray wavelengths, but not seen in
\Halpha.

Detection criteria can be formulated in order to predict whether a
pulsar should produce a bow shock nebula visible in \Halpha.
Considering the average flux per detector pixel, the ratio of the
\Halpha\ flux to the angular size of the nebula, $F_\alpha / \Delta
\Omega$, must exceed a (detector dependent) threshold:
\begin{equation}
\frac{F_\alpha}{N_{\rm pixel}} = \frac{F_\alpha}{\Delta \Omega
/\Delta_{\rm CCD}} > S_{\rm thresh}.
\end{equation} 
From \Eref{falpha}, this implies that $X \nH v_{\ast} / \gHa$
must exceed a limiting value, a result independent of the spindown
flux $\dot{E}/D^2$. However, a criterion based on the average flux
does not account for brightness variations within the nebula, which
can be substantial (as seen, for example, in the Guitar Nebula,
discussed in \Sref{GN}).  The optimal detection method is to block
average all pixels which contain nebular emission, in which case 
(assuming uncorrelated random noise in each pixel) the
average flux per detector pixel only needs to exceed $S_{\rm
thresh}/\sqrt{N_{\rm pixel}}$ for a detection:
\begin{equation}
\frac{F_\alpha}{N_{\rm pixel}} >
\frac{S_{\rm thresh}}{\sqrt{N_{\rm pixel}}}.
\end{equation} 
It is worth noting, however, that the optimal averaging size may not
be known in advance, and a computationally intensive search may be
required.  For optimal detection, again using \Eref{falpha} and
substituting for the angular size of the nebula,  
\begin{equation}
\frac{X \nH^{1/2}}{\gHa} \frac{\dot{E}^{1/2} v_\ast^\beta}{D}
> \kappa \, S_{\rm thresh}, 
\label{Eqn:detect}
\end{equation}
where $\kappa$ encapsulates various constants and numerical factors.
As expected, \Halpha\ bow shock nebulae are more likely to be detected
for pulsars with larger spindown flux and higher velocities, located
in regions with higher neutral fractions and lower extinction.

The detection criterion formulated in \Eref{detect} can be recast in
terms of the minimum NS velocity required to produce a detectable
bowshock, given the spindown flux and the ambient ISM. For
concreteness, we assume a telescope sensitivity comparable to the 5-m
Hale Telescope at Palomar, as used to detect the Guitar Nebula
(\Sref{GN}), and $\beta = 1$ as before. Then, for $D$ in kpc, \Edot\
in units of 10$^{33}$ erg \persec, and $\nH$ in cm$^{-3}$,
\begin{equation}
V_{\rm min} = 14\;{\rm km \; s}^{-1} \frac{\gHa}{X \nH^{1/2}} 
\left(\frac{\dot{E}_{33}}{D^2_{\rm kpc}}\right)^{-1/2}.
\label{Eqn:detectvel}
\end{equation}

The relationship derived for the minimum required velocity is plotted
in \Fref{detect}, along with transverse velocities for all pulsars
with measured proper motions. All known bow shock nebulae are marked
on the plot: their locations confirm the applicability of the
detection criterion developed above.  As mentioned above, the X-ray
and radio-detected bow shock nebulae which are not detected in
\Halpha\ provide upper limits on the neutral fraction present in the
ambient medium.  The plot also identifies future candidates for deep
searches for bow shock nebulae, although the uncertainties in the
distance and velocity estimates for most pulsars and the large range
of plausible values for $X \nH^{1/2} / \gHa$ are likely to limit its
usefulness as a strong predicting tool.

\begin{figure}[hf]
\epsscale{0.85}
\plotone{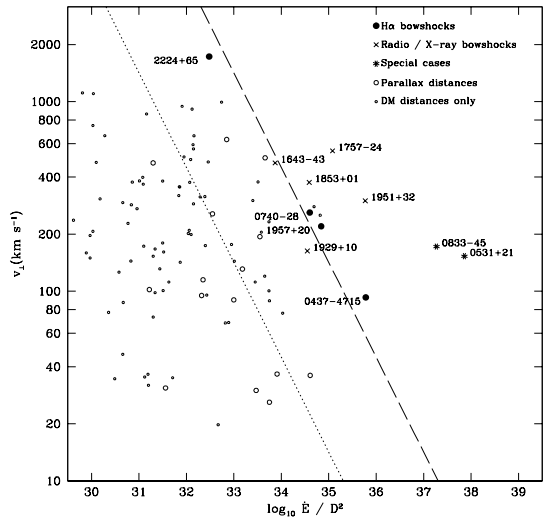}
\caption{Detectability of bow shock nebulae. Transverse velocities are
plotted against the spindown flux $\dot{E}/D^2$ for all pulsars with
published proper motions.  The velocity and distance estimates are
significantly better for pulsars with parallax distances, which are
plotted with larger open circles. Lines are overplotted for the
detection criterion in \Eref{detectvel}, for $X \nH^{1/2} / g_\alpha =
0.01$ (dashed) and 0.1 (dotted).  All known \Halpha\ and radio / X-ray
bow shock nebulae are identified on the plot.  The Crab (B0531+21) and
Vela (B0833$-$45) pulsars are not detected as bow shock nebulae
because of the negligible neutral fraction in their local ISM. }

\label{Fig:detect}
\end{figure}

\section{MULTI-WAVELENGTH OBSERVATIONS OF THE GUITAR NEBULA}\label{Sec:GN}

The visually spectacular Guitar Nebula is a bow shock nebula produced
by the otherwise unexceptional pulsar B2224+65 ($P=0.68$ s,
$\dot{P}=9.7 \times 10^{-15}$ s~\persec). The spindown luminosity of
the pulsar ($\dot{E}=10^{33.08}$ erg~\persec, assuming a moment of
inertia of $10^{45}$ gm cm$^2$) is significantly lower than the \Edot\
of other pulsars with known \Halpha\ bow shocks. However, B2224+65 has
an extremely large space velocity, which evidently compensates for the
low \Edot\ in producing a visible bow shock nebula.  Its proper motion
$\mu = 182 \pm 3$ mas yr$^{-1}$, at a position angle $52.1\arcdeg \pm
0.9\arcdeg$ \citep{HLA93}.  The dispersion measure (DM = 36.16 $\pm$
0.05~pc~cm$^{-3}$) of the pulsar, combined with a model for the
Galactic electron density distribution \citep{TC93} yields a distance
$D \approx 2.0 \pm 0.5$~kpc, at which the transverse velocity
$v_{\perp} = D \mu \approx (1.7 \pm 0.4) \times 10^{3}$~km~\persec, the
largest known velocity among radio pulsars.

It is evident from the discovery images that there may be a
significant amount of ionized material in the vicinity of the pulsar,
biasing the DM upward, and hence the distance may be overestimated.
However, even at a distance of 1~kpc, the pulsar velocity ($\gtrsim
850$ km~\persec) is significantly higher than the mean population
velocity of $\sim 450$ km~\persec\ estimated by \citet{LL94}.  Recent
studies \citep{ACC01,CC98} prefer a bimodal velocity distribution,
with a significant portion ($\sim 15\%$) of pulsars having a space
velocity $> 1000$ km~\persec. B2224+65 is probably at the high end of
this distribution, and the bow shock nebula is visible in spite of the
small \Edot\ due to the large space velocity of the pulsar.

\subsection{Ground-based Optical Observations}

\begin{figure*}[htf]
\epsscale{1.5}
\plotone{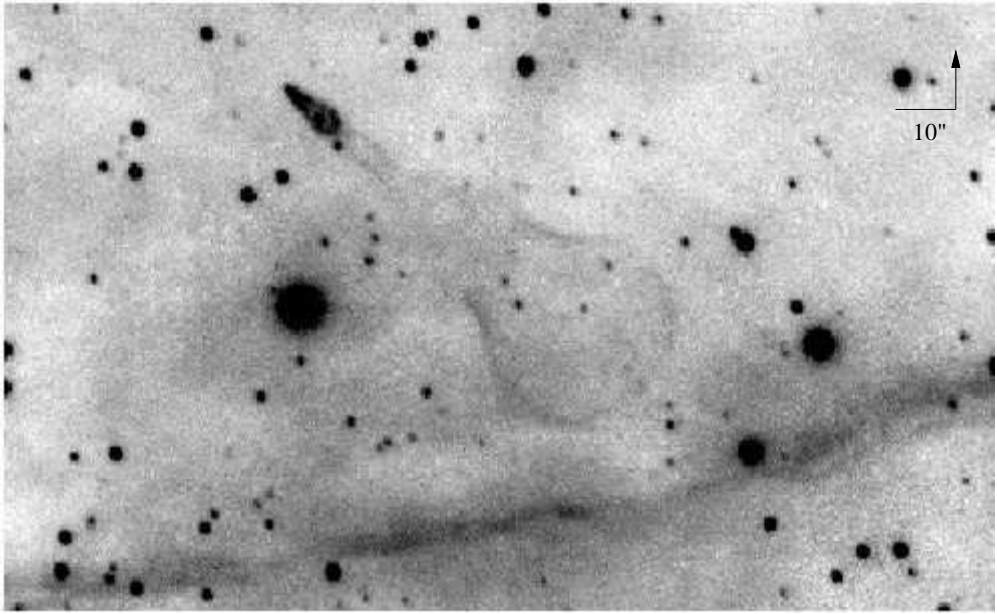}
\caption{The \GN\ in \Halpha\ (20 \AA\ filter at 6564 \AA), imaged
with the 5-m Hale Telescope at Palomar Observatory in mid-1995. North
is up and east is to the left, as indicated, and the L has arms
10\arcsec\ in size.}
\label{Fig:palomar}
\end{figure*}

In an ongoing program at Palomar Observatory, the \GN\ has been imaged
regularly at the 5-m Hale Telescope using the Carnegie Observatories
Spectroscopic Multislit and Imaging Camera \citep[COSMIC;][]{cosmic},
mounted at the f/3.5 prime focus, in reimaging mode.  A narrow-band
\Halpha\ filter (20~\AA\ at 6564~\AA) was used.  The typical seeing
was $\sim 0.9\arcsec$--$1.4\arcsec$, significantly larger than the
CCD pixel scale (0.4\arcsec).  The CCD data were processed in
standard ways with IRAF tasks. Data obtained in 1995 were combined
(9000 s total integration time) to produce the representative image
presented in Fig.~\ref{Fig:palomar}.
 
The Palomar image of the \GN\ shows a remarkable structure consisting
of a faint, limb-brightened ``guitar body'' which is approximately
symmetric, with an elongated neck and a bright head, the tip of which
coincides with the location of the pulsar. The axis of the nebula
matches the position angle of the pulsar proper motion ($52.1\arcdeg
\pm 0.9\arcdeg$), as expected. Differential Galactic rotation causes
only a small effect ($<3\%$ of the transverse velocity) and is
therefore neglected. 

The body of the guitar appears to be composed of two closed-off
bubbles. This could be explained by the projection of an open
structure onto the plane of the sky, or by episodic variation in
\Edot, which is not inconceivable given that B2224+65 has been known
to glitch. The complex shape may alternatively be explained by the
penetration of neutral ISM atoms into the space occupied by the pulsar
wind, as suggested by \citet{BB01}.  However, the simplest explanation
is the existence of density variations in the ambient medium.  The
observed closed bubble structures can be produced by density gradients
oriented roughly parallel to the NS transverse velocity.  The nebula
brightens where it narrows between the two bubbles, as expected in a
higher density region. The extremely bright head is then created by
the NS entering a region with much higher ($\sim 10 \times$) density,
or a significantly higher neutral fraction.  The distinct variations
in brightness along the nebula, the observed departures from symmetry
on small scales and the existence of large-scale filamentary
structures (visible at the bottom of \Fref{palomar}) can all be
explained within this scenario, although other explanations cannot be
ruled out.

\subsection{Radio Interferometric Observations}

The \GN\ was observed at radio wavelengths with the NRAO Very Large
Array (VLA). In order to get comparable resolutions over a range of
frequencies, different VLA configurations were used: B-array for
1.4~GHz and 4.8~GHz, and C-array for 8.4~GHz observations. About 3
hours of integration time were obtained at each frequency, though the
8.4~GHz observation was plagued by poor weather conditions and did not
yield as much usable data as the other frequencies. The observation
parameters are summarized in \Tref{VLA}.

The data were deconvolved using the CLEAN algorithm \citep{clean}, as
implemented in AIPS, the Astronomical Image Processing System, in
order to remove the sidelobes of strong sources within the primary
beam. Wide-field images were produced at each frequency with map noise
ranging from 23 $\mu$Jy to 28 $\mu$Jy, as detailed in the last column
of \Tref{VLA}. 
 
\begin{deluxetable}{cccccc}
\tablecolumns{6}
\tablewidth{0pc} 
\tablecaption{Multi-array Radio Observations at the VLA\label{Table:VLA}}
\tablehead{ 
\colhead{$\nu_{\rm obs}$} & \colhead{VLA} & \colhead{Beam} & 
\colhead{T$_{\rm int}$} & \colhead{$\Delta\nu$} & \colhead{$\sigma$ (RMS)} \\
\colhead{(GHz)} & \colhead{Array} & \colhead{(\arcsec)} &
\colhead{(hr)}  & \colhead{(MHz)} & \colhead{($\mu$Jy/beam)}
}
\startdata
1.4 & B & 3.8 & 2.9 & 50 & 27 \\
4.9 & B & 1.2 & 2.9 & 50 & 23 \\
8.4 & C & 2.2 & 2.4 & 100 & 28

\enddata 
\end{deluxetable} 
 
The radio pulsar B2224+65 is detected at 1.4~GHz as a point-source
with a flux $\sim$ 3.2 mJy. It is undetected at the higher
frequencies. No extended features related to the \GN\ or the optical
filamentary structures are observed. We derive an upper limit ($\sim 5
\sigma$) of 0.1~mJy beam$^{-1}$ on any synchrotron emission, either
from the relativistic pulsar wind, or from the shock accelerated
particles in the nebula, at wavelengths between 1.4 and 8.4~GHz with
1--4\arcsec\ beam sizes.  The non-detections thus rule out any
plerionic component to the pulsar or any non-thermal emission from the
\GN\ over the observed range of wavelengths and resolutions.  The data
were also smoothed by tapering in the $U-V$ domain: although the
possibility of faint extended emission on much larger scales cannot be
excluded, no significant flux was detected from the Guitar Nebula
after smoothing to 2--8\arcsec\ beam sizes, comparable to the size of
the bright part of the nebula in \Halpha.

The derived upper limit on plerionic radio emission is more stringent 
than those derived by \citet{GSF+00} for other pulsars, and the
non-detection is consistent with their conclusion that older pulsars
may become less efficient at producing synchrotron emission with
increasing age. 

\subsection{X-ray and Infra-Red Observations}

X-ray observations of the \GN\ using the ROSAT High Resolution Imager
have been described by \citet{RCY97}. Weak soft X-ray emission is
detected using an aperture mask matched to the optical nebula, but the
detection is significant at only $\sim 4 \sigma$. Follow-up
observations using Chandra have recently been obtained, and will be
described elsewhere.  

Narrow-band infrared observations with PFIRCAM at the Hale 5-m
telescope at Palomar Observatory resulted in non-detections, which
complement the radio observations in limiting any thermal or
non-thermal emission from the \GN.  Upper limits on the infrared flux
will be published in future work.

\section{HUBBLE SPACE TELESCOPE OBSERVATIONS AND MODELING}\label{Sec:HST}

The head of the Guitar Nebula is not well resolved by ground-based
observations, and the expected scale of the shocked region where the
NS wind meets the ambient medium ($\sim 0.05\arcsec$) cannot be probed
from the ground. HST observations are required, followed by model
fitting to obtain the bow shock parameters.

\begin{figure*}[htf]
\epsscale{1.8}
\plotone{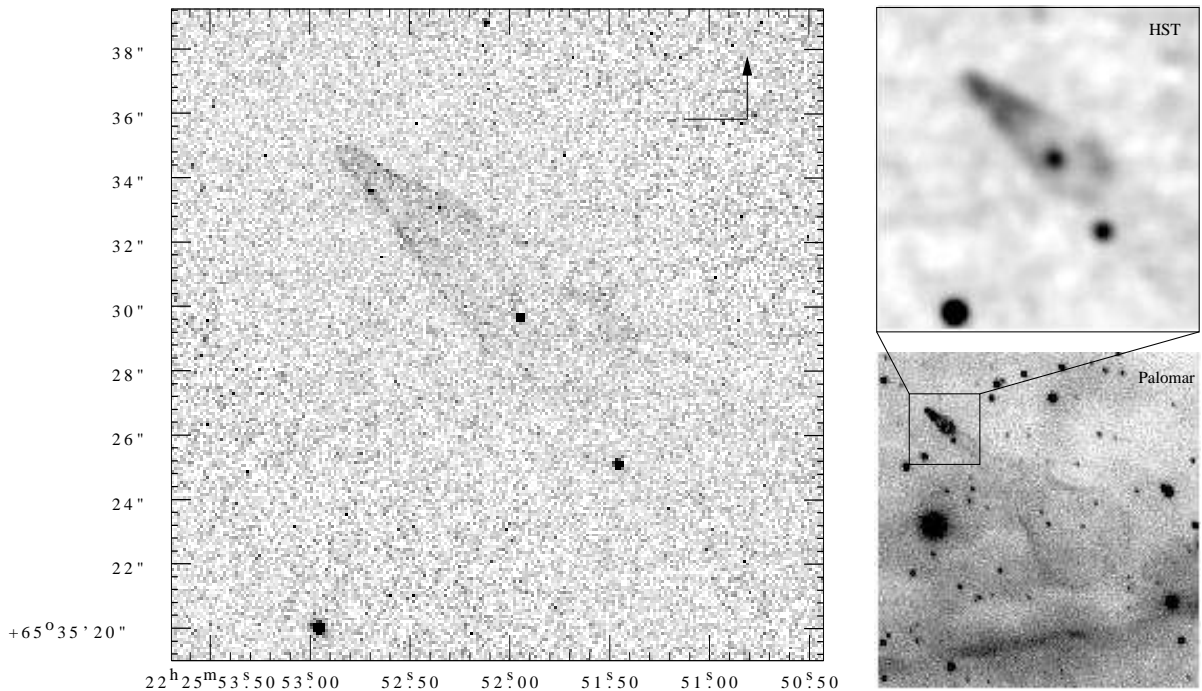}
\caption{The head of the \GN, imaged with the HST Planetary Camera
(7200 seconds, 22 \AA\ filter at 6564 \AA) in late 1994.  The left
panel shows a section of the image with a hyperbolic-sine stretch
applied to the grayscale.  The lower right panel marks the region
covered by the image section, superimposed on the near-contemporaneous
Palomar image (mid 1995). The upper right panel shows the HST image
convolved with the typical seeing disk ($\sim 1\arcsec$ FWHM) at
Palomar, which makes the bright head and the neck of the Guitar
visible.  For reference, the three bright stars in the HST images are
also visible on the Palomar image (\Fref{palomar}). (Image degraded to
fit astro-ph size limits)}
\label{Fig:hst}
\end{figure*}

\subsection{HST Observations}

High-resolution HST observations were obtained in 1994 December, using
the Wide Field and Planetary Camera 2 \citep[WFPC2;][]{wfpc2}.  The
bright head of the nebula was centered on the Planetary Camera (PC),
which provides 0.0455\arcsec\ resolution.  The observations include 6
dithered exposures of 1200 seconds each, using filter F656N (22 \AA\
at 6564 \AA), comparable to the filter used at Palomar. The images
were combined using Variable-Pixel Linear Reconstruction (or
`Drizzling'), as described by \citet{drizzle}. This includes cosmic
ray rejection and partially compensates for the undersampling of the
HST point-spread function by the 0.0996\arcsec\ pixels of the WF chips
on WFPC2.

A section of the PC image is presented in \Fref{hst}: for
reference, the three stars visible in the frame are also seen in the
Palomar image (\Fref{palomar}).  In spite of the limited
signal-to-noise ratio (S/N) of the image, it is apparent that the
bright head of the \GN, unresolved in the Palomar images, is resolved
by the PC. The bow shock is limb-brightened, as observed in the
extended Guitar body observed at Palomar.  The bright head and the
neck of the Guitar become visible when the PC image is smoothed down
to the typical seeing at Palomar, as shown in the upper right panel of
\Fref{hst}.  The extended body of the Guitar and the nearby
filamentary structures are also detected in the WF images after
smoothing (not shown here). 

Along with the narrow-band images, broad-band images were also
obtained with WFPC2, with 4 dithered exposures of 180 seconds each
using filter F675W (867 \AA\ at 6717 \AA). The combined image (720 s)
shows no significant emission from the nebula. As expected, no optical
counterpart is detected for the radio pulsar, down to a limiting
magnitude of $R = 25.8$ ($5 \sigma$) on the broad-band PC image.

\subsection{Evolution of the Guitar Nebula}

As discussed above, the simplest explanation for the characteristic
shape of the \GN\ is the propagation of the NS through a
non-uniform ambient medium, in which case the shape of the nebula
provides information about the density of the medium encountered in
the past. 

\Fref{sym} shows the symmetric and anti-symmetric components of the
head of the Guitar Nebula, which were obtained by mirroring a smoothed
version of the PC image about the axis of the nebula and taking the
sum and the difference of the two.  The axis was found by using the
pulsar proper motion as a starting point, and then fitting for the bow
shock position angle, as described below.  The symmetric component of
the image shows significant changes in the opening angle of the nebula
at 2\arcsec\ scales, which are also visible, though less prominent, in
the unsmoothed HST image (\Fref{hst}).  It appears that the opening
angle has narrowed recently ($\lesssim 10$ yr, between 1 and 2\arcsec\ 
downstream from the tip), probably as the NS encountered a gradient of
increasing density.

\begin{figure}[ht]
\epsscale{0.95}
\plotone{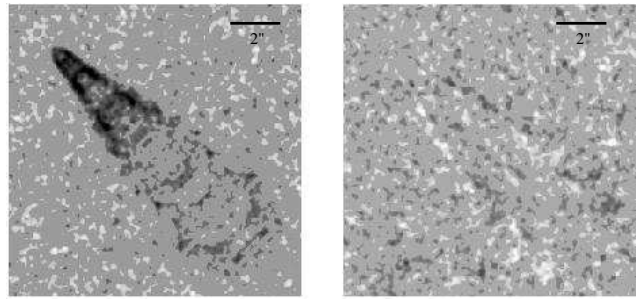}
\caption{Symmetric and anti-symmetric components of the head of the
Guitar Nebula, obtained by mirroring the head of the nebula about its
axis.  The image has been enhanced to bring out detail by assigning
the mean amplitude to all pixels with amplitudes within $\pm 1 \sigma$
of the image mean. The range of the greyscale is the same for both
panels.  While the tip of the head is almost perfectly symmetric, the
limbs of the nebula are visible in the anti-symmetric image at the
$\pm 2 \sigma$ level.}
\label{Fig:sym}
\end{figure}

There is also a small but significant departure from bilateral
symmetry along the head of the Guitar Nebula, as illustrated by the
image of the anti-symmetric component.  The tip of the head is almost
perfectly symmetric, and disappears in the anti-symmetric image, but
the north-west and south-east limbs of the nebula deviate from the
axis of symmetry at the $2 \sigma$ level, consistent with a density
gradient in the ambient medium which is not aligned with the NS
velocity.

In both the smoothed HST image (\Fref{hst}) and the image of the
symmetric component, the head of the nebula appears to have a
closed-off structure, reminiscent of the rounded end of the ``guitar
body'' seen on the Palomar image (\Fref{palomar}). Once again, while
various other explanations (including a time-variable \Edot) cannot be
ruled out, the simplest explanation involves the existence of a
density gradient, which confines the expansion of the head of the
nebula. This explains both the brightening of the head of the nebula
and the faint tubular ``neck'' of the guitar.

As part of an ongoing effort to monitor the evolution of the nebula,
several images have been obtained at the 5-m Hale Telescope at Palomar
Observatory over the past 7 years. Additionally, we expect to obtain
another HST observation at higher S/N in late 2001, at which time the
NS will have moved $\sim 1\arcsec$ from the previous position. This
will allow a detailed study of the dynamic evolution of the head of
the nebula and the ISM in the neighborhood of the NS, or
alternatively, more exotic scenarios with time-variable \Edot.

\subsection{Modeling the Bow Shock}

The lack of an optical counterpart to PSR~B2224+65 makes it difficult
to obtain bow shock parameters like the stand-off angle $\theta_0$ by
inspection of the image. Model fitting must be used in order to
extract $\theta_0$ from the HST image, which resolves the stand-off
region at the head of the bow shock.

Adopting the simple momentum-conserving description from \citet{W96},
the shape of the bow shock can be expressed in polar coordinates
($R,\theta$) with the NS at the origin and traveling along the
$\hat{z}$ axis:
\begin{equation}
R(\theta) = \frac{\sqrt{3} R_0}{\sin\theta} 
	    \left(1-\frac{\theta}{\tan\theta}\right)^{1/2}.
\label{Eqn:shape} 
\end{equation}

This expression describes the shape of a thin shell bow shock provided
that the shocked gas undergoes rapid radiative cooling.  If the
cooling is not rapid enough, thermal pressure should result in a
somewhat wider tail downstream along the bow shock.  The
momentum-conserving description also assumes mixing in the shocked
layer between the ambient medium and the relativistic wind, whereas a
laminar flow may be a more accurate description \citep{BB01}.  The
model also assumes an ambient medium with uniform density, which, we
have argued, is not the case here.  However, the present work is
limited to extracting basic bow shock parameters from an image with
poor S/N.  To work around these limitations, the model fitting
procedure was restricted to within 2.6\arcsec\ of the tip of the bright
head, where the opening angle does not change much, the surface is
relatively smooth, and the behavior of the tail downstream is not
relevant.

The thin shell bow shock described by \Eref{shape} can be
scaled and projected to account for viewing geometry. For each set of
scale and geometry parameters, the projected surface density is
calculated by integrating the density through the bow shock along each
line of sight. The \Halpha\ flux from the nebula is proportional
to the projected surface density, accounting for the observed
limb-brightened appearance.

A bow shock nebula described by this model can be parameterized by two
angles, the inclination angle of $\hat{z}$ to the LOS ($i = 0$ for
motion towards the observer, and $i = \pi/2$ for motion in the plane
of the sky), and a position angle on the sky, measured east of
north. Additional parameters are the thickness of the shocked layer
which emits in \Halpha\ as a fraction of the stand-off radius, a
constant numerical factor which relates the \Halpha\ flux to the
projected surface density, and a dimensionless scale parameter,
\begin{equation}
S = D^{-2}_{\rm kpc}\, \nA^{-1/2} (\dot{E}/\dot{E}_{I_{45}})^{1/2},
\end{equation}
which accounts for the apparent change in size with distance $D$, the
number density of the ambient medium $\nA$, and the ratio of the
actual NS luminosity to the calculated spindown luminosity.

Models were generated over a grid of parameter values for position
angle (43\arcdeg--53\arcdeg), scale parameter $S$ (1.0--6.0),
inclination to the LOS $i$ (15\arcdeg--90\arcdeg) and the thickness
of the \Halpha -emitting layer (5\%--25\% of the stand off radius).
These models were compared to the part of the \GN\ least affected by
the non-uniformity of the ambient medium, within $\sim 2.6$\arcsec\ of
nebula tip in \Fref{hst}, which covers $\sim 6,600$ pixels on the PC
image.  Best fit parameters were obtained using least-squares
minimization of the difference between the data ($D_k$) and the model
($M_k$) over the image ($1 \le k \le N$, where $N$ is the number of
pixels),
\begin{equation}
\chi^2 = \sum_{k=1}^{N} \sigma^{-2} (D_k-\alpha M_k)^2,
\end{equation}
where $\alpha$ is the numerical factor relating the \Halpha\ flux to
the projected surface density, calculated using a matched filtering
approach: 
\begin{equation}
\alpha = \frac{\sum_{k=1}^{N} D_k M_k}{\sum_{k=1}^{N} M_k^2}.
\end{equation}

The model fitting procedure yields a good estimate for the stand-off
angle $\theta_0$ separating the shock front and the NS.  For the
entire range of plausible models, we obtain $\theta_0 = 0.06\arcsec
\pm 0.02\arcsec$.  The position angle is also determined to be
$48\arcdeg \pm 2\arcdeg$, in agreement with the pulsar proper motion
to within the errors.  The measured stand-off angle is used to test
scaling laws for bow shock nebulae in the next section; we note here
that the measured $\theta_0$ can be used in \Eref{numtheta} to
constrain the distance to B2224+65:
\begin{equation}
D_{\rm kpc} = 0.75 \frac{\sin i}{n_A^{1/4}}
\label{Eqn:D}
\end{equation}

However, model fitting was inconclusive for the other parameters,
especially the inclination angle $i$. A minimum was found in the
$\chi^2$ surface at $i=90\arcdeg$, with reduced $\chi^2 = 1.20$,
compared to a reduced $\chi^2 = 1.30$ for a model consisting of a flat
background only.  However, the surface is not simple, and shows
another secondary minimum at $i \sim 30\arcdeg$ (reduced $\chi^2 =
1.21$).  The corresponding values of the scale factor $S$ are 5.0 and
2.5 respectively.  Physically, $i \sim 90\arcdeg$ is preferred, since
a low inclination angle implies an extremely high three-dimensional
velocity. The primary reason for the poor fit is the low S/N in the
image.  Observations scheduled for late 2001 with the HST WFPC2 will
have a significantly longer integration time, and hence a better S/N,
which is expected to allow a more discriminating model fit.

\begin{deluxetable}{lcccccl}
\tablecolumns{7}
\tablewidth{0pc} 
\tablecaption{\Halpha\ Bow Shock Parameters\label{Table:scale-Ha}}
\tablehead{ 
\colhead{Pulsar} & \colhead{D} & \colhead{$\mu$} & 
\colhead{$\log_{10} \dot{E}$} & \colhead{$\theta_{0}$} & 
\colhead{$F_{\alpha}$} & \colhead{Refs.} \\
\colhead{} & \colhead{(kpc)} & \colhead{(mas yr$^{-1}$)} & 
\colhead{(erg s$^{-1}$)} & \colhead{(\arcsec)} & 
\colhead{(cm$^{-2}$ s$^{-1}$)} & \colhead{}
}
\startdata
J0437$-$4715 
             & $0.139 \pm 0.003$ & $140.892 \pm 0.006$ &  34.07 &
	 $10.0 \pm 1.5$ & $ 2.5 \times 10^{-3}$ & 1,2 \\
B0740$-$28 & $1.9 \pm 0.5$ & $29.0 \pm 0.9$ &  35.16 & 
	$1.0 \pm 0.2$ &  $5 \times 10^{-5}$  & 3 \\
B1957+20 & $1.5 \pm 0.4$ & $30.4 \pm 0.6$  &  35.20 &
	  $4.0 \pm 0.6$ & $1.09 \times 10^{-4}$& 4 \\
B2224+65 & $2.0 \pm 0.5$ & $182 \pm 3$ &  33.08 & 
         $0.06 \pm 0.02$ & $1.1 \times 10^{-3}$ & 5,6 \\
\\
J1856.5$-$3754 
         & $0.14 \pm 0.04$ & $332.0 \pm 0.8$ & $32.9 \pm 0.3$ & 
	 $1.0 \pm 0.2$ & $2 \times 10^{-5}$& 7,8,\dag \\
\enddata
\tablerefs{(1) \citet{BBM+95}; (2) \citet{VBB+01}; (3) \citet{JSG02};
(4) \citet{KH88}; (5) \citet{CRL93}; (6) this work; (7) \citet{KVA02}; 
(8) \citet{VK01}. (\dag: Not detected as a radio or X-ray pulsar.)}
\end{deluxetable}

The narrowing of the opening angle of the bow shock and the departure
from bilateral symmetry, which are illustrated in \Fref{sym},
may be modeled by a density gradient in the ambient medium which is
not aligned with the NS velocity vector. \citet{W00} provides a more
general treatment than the current one, and this may be used for
future modeling, along with a full numerical solution.  However, the
current image S/N does not justify more complex models with additional
free parameters.

\section{TESTS OF SCALING LAWS}\label{Sec:test}

\subsection{\Halpha\ Bow Shocks}

The bow shock parameters obtained for the Guitar Nebula can be used,
in conjunction with parameters measured for other known bow shock
nebulae, to test the scaling laws derived in \Sref{scale}.  Besides
the Guitar Nebula, bow shocks have been detected in \Halpha\ for the
radio pulsars B1957+20 \citep{KH88}, B0437$-$4715 \citep{BBM+95} and
B0740$-$28 \citep{JSG02}. In each case, the spectra are dominated by
Hydrogen Balmer lines.  While the proper motion and \Edot\ of these
pulsars are well measured, the distances are obtained from the pulsar
DM along with a model for the Galactic electron density \citep{TC93},
and are therefore relatively poorly constrained.  The exception to
this is PSR B0437$-$4715, which has an extremely precise parallax
measurement from timing observations \citep{VBB+01}.

The observable parameters for these pulsars are summarized in
\Tref{scale-Ha}, with {\em ad hoc} 15--20\% errors assigned to
$\theta_0$ measurements except for the Guitar Nebula.  It is worth
noting that \Edot\ spans more than two orders of magnitude for these
four pulsars, and the proper motions and distance estimates span an
order of magnitude as well.  The scaling of the stand-off angle for
\Halpha\ bow shocks is demonstrated in \Fref{scale-Ha}, where the
observed stand-off angle is plotted against the other observable
quantities, and lines of constant $n_A$ are overplotted for
$i=90\arcdeg$ and $45\arcdeg$.  Due to projection along the LOS,
$\theta_0 \le R_0/D$.  Thus, for each observed bow shock nebula, the
lines of constant $\nA$ for $i=90\arcdeg$ place an upper limit on the
density of the ambient medium, independent of the actual inclination
to the LOS.  For example, for the Guitar Nebula, the nominal ambient
density is constrained to be $\nA \lesssim 0.015$~cm$^{-3}$, which
corresponds to $\nH \lesssim 0.011$~cm$^{-3}$ for a Helium fraction of
0.27 by mass (cosmic abundance) in the ISM (i.e.\ $\gammaH = 1.37$).

The scaling laws derived for \Halpha\ bow shocks associated with radio
pulsars can also be generalized to radio-quiet NS, under the
assumption that these NS have similar relativistic winds.
RX~J1856.5$-$3754 is the nearest known NS, with a published HST
parallax distance of $61 \pm 9$~pc \citep{W01}.  This distance has
been called into question by \citet{KVA02}, who reanalyze the same
data to obtain a parallax distance $= 140 \pm 40$~pc, which we adopt
here.  While the lack of radio pulsations can be explained by
postulating a mis-directed pulsar, the absence of detected X-ray
pulsations from thermal hot spots \citep{RGS02} is more puzzling.  An
associated cometary \Halpha\ nebula has been detected by \citet{VK01},
who describe it as being consistent with either a bow shock or an
ionization nebula. Assuming that it is a bow shock (which is the
simpler of the two possibilities, especially in the light of the
revised distance) and using a formalism similar to this work,
\citet{VK01} estimate $\dot{E} \simeq 6 \times 10^{31}$~erg~\persec,
which increases to $\simeq 8 \times 10^{32}$~erg~\persec\ for the
revised distance.  From \Fref{scale-Ha}, $\nA \lesssim 1$~cm$^{-3}$,
consistent with the value derived by \citet{VK01}. 

It is worth emphasizing the diversity of the bow shock nebulae in
\Fref{scale-Ha}. B1957+20 and J0437$-$4715 are millisecond radio
pulsars while RX~J1856.5$-$3754 is a radio-quiet NS, and all three
produce canonical bow shock nebulae. On the other hand, B2224+65 and
B0740$-$28 are ordinary radio pulsars but the bow shock nebulae they
produce do not have simple shapes: the former produces the Guitar
Nebula while the latter produces a nebula shaped like a key hole.  
Faint X-ray emission has been detected by ROSAT for B2224+65
\citep{RCY97} and B1957+20 \citep{FBGB92}, although in neither case is
the emission definitively associated with the bow shock nebula.
In spite of this diversity, the stand-off angle, at least, is well
parameterized by a few simple physical quantities, as expected from
the underlying physics.

\begin{figure}[hf]
\plotone{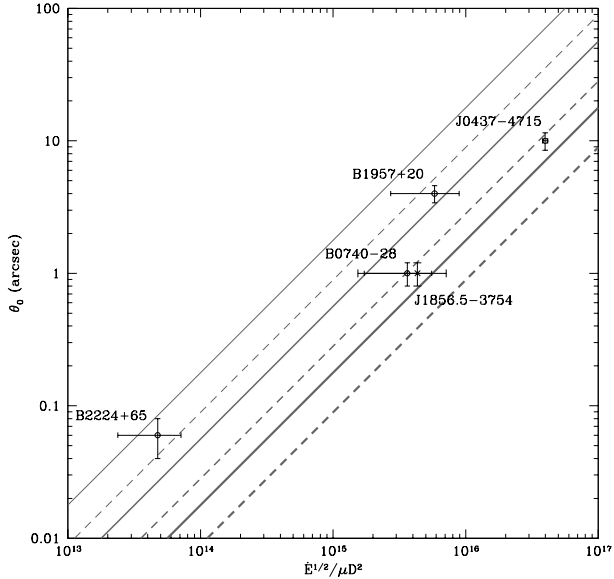}
\caption{Scaling of the stand-off angle for \Halpha\ bow shock nebulae
with other observable parameters, as listed in \Tref{scale-Ha}.
Circles mark radio pulsars and the cross marks RX~J1856.5$-$3754, a
radio-quiet NS.  Lines of constant number density $\nA =$ 0.01, 0.1,
1.0~cm$^{-3}$ are overplotted for inclination to the line of sight $i
= 90\arcdeg$ (thinnest to thickest solid lines) and $i = 45\arcdeg$
(thinnest to thickest dashed lines).}
\label{Fig:scale-Ha}
\end{figure}

The total \Halpha\ flux from a bow shock nebula also scales with a
combination of measured, observable and unknown parameters, as
described by Eqns.~\ref{Eqn:f1} and \ref{Eqn:f2}.  While \Edot, $\mu$,
$D$ and $\theta_0$ are accessible to observation, the density $\nH$ is
unknown, and few {\em a priori} constraints exist on the neutral
fraction $X$. Additionally, the scaling is a function of how the size
of the bow shock depends on the velocity, parameterized as $\Delta
\Omega \propto v_{\ast}^{2\beta}$, and we have argued that $\beta=1$.

The five bow shock nebulae with published \Halpha\ flux values
($F_{\alpha}$) are plotted in \Fref{scale-flux} against two different
combinations of observable and measured parameters, with $\beta=1$.
While a correlation is apparent in the plot of $F_{\alpha}$ against
$\theta_0^2 \mu^3 D^3$, it is worth noting that differences in the
density and neutral fraction in the environment of each object and the
different extinctions along each LOS are not accounted for in this
plot, and neither is the projection effect for the velocity ($\sin^3
i$), which differs for each bow shock.  On the other hand, for
$\beta=0$ (nebula size independent of velocity) and $\beta=0.5$, the
points appear scattered with no discernible trend, favoring $\beta=1$
and $\Delta \Omega \propto v_{\ast}^2$.  The two combinations of
observables ($\theta_0^2 \mu^3 D^3$ and $\dot{E}\mu / D$) are related
to each other through \Eref{theta0}, with differences in the medium
density accounting for the relative scatter between the two plots.

\begin{figure}[htf]
\epsscale{1.0}
\plotone{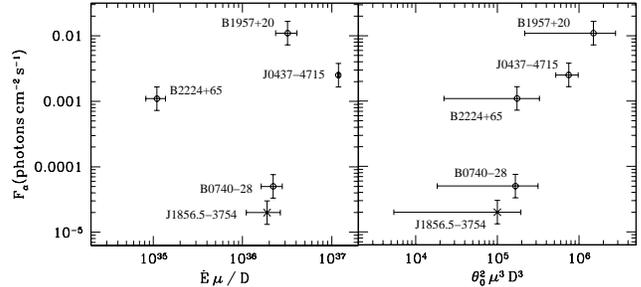}
\caption{Scaling of \Halpha\ flux from bow shock nebulae with two
different combinations of observable parameters, for \Edot\ in
erg~\persec, $\theta_0$ in arcsec, $\mu$ in mas~yr$^{-1}$ and $D$ in
kpc. Circles mark radio pulsars and the cross marks RX~J1856.5$-$3754,
a radio-quiet NS.  The effects of the different inclinations to the
LOS and the differing extinctions, densities and neutral fractions are
not accounted for in these plots, and are responsible for the relative
scatter between the two panels.}
\label{Fig:scale-flux}
\end{figure}
 
We note that there are two independent observed quantities ($\theta_0$
and $F_\alpha$) which depend on the measurable values for \Edot, $\mu$
and $D$ (as derived in \Sref{scale}). In principle, these can be
combined to jointly constrain the unknown values of the density $\nH$,
neutral fraction $X$ and the inclination to the LOS. While current
observations still lack the precision to produce meaningful
constraints on all the unknown parameters at the same time, a parallax
distance to PSR B2224+65 or a more precise measurement of the bow
shock parameters for J0437$-$4715, for example, can be exploited in
the future to obtain such constraints.

\subsection{Bow Shocks at Radio/X-ray Wavelengths}

As mentioned in \Sref{scale}.3, NS bow shocks consist
of two shocked regions: an outer shock, where the ambient ISM
undergoes collisional excitation and may emit in \Halpha, and an inner
shocked region enclosing the relativistic NS wind, which may emit
synchrotron radiation.  No bow shocks have currently been detected in
both \Halpha\ and synchrotron radiation at radio wavelengths, although
a faint extended X-ray counterpart may exist for the \GN.  It is
possible that synchrotron-bright nebulae are observed only for younger
NS, and radiation from the young NS or the parent supernova
pre-ionizes the ISM, suppressing \Halpha\ emission, or alternatively,
that these younger NS are more likely to be found in superbubbles
which have insignificant neutral fractions. Future observations
should resolve this issue.  Meanwhile, it is satisfying to recognize
that the scaling argument for the stand-off radius developed for
\Halpha\ bow shocks can be applied to radio and X-ray detected
synchrotron nebulae as well, in cases where a cometary morphology and
a known radio pulsar indicate that a bow shock mechanism is at work.

There are several such radio pulsars which have associated plerionic
bow shock nebulae at radio wavelengths, but not in \Halpha: for
example, B1951+32 \citep{HK88}, B1757$-$24 \citep[the Duck,][]{FK91},
B1853+01 \citep{FGGD96} and B1643$-$43 \citep{GFGV01}. Each of these
has a clear association with a supernova remnant, attesting to the
relative youth of the object.  The best available parameters for these
pulsars and the stand-off angles of the associated bow shocks (either
quoted or estimated from the radio images) are listed in
\Tref{scale-radio}.

\begin{deluxetable}{lcccccl}
\tablecolumns{7}
\tablewidth{0pc} 
\tablecaption{Radio and X-ray Bow Shock Parameters \label{Table:scale-radio}}
\tablehead{ 
\colhead{NS} & \colhead{SNR} & \colhead{D} & \colhead{V} & 
\colhead{$\log_{10} \dot{E}$} & \colhead{$\theta_{0}$} & \colhead{Refs.} \\
\colhead{} & \colhead{} & \colhead{(kpc)} & \colhead{(km s$^{-1}$)} & 
\colhead{(erg s$^{-1}$)} & \colhead{(\arcsec)} & \colhead{}
}
\startdata
  B1643$-$43 & G341.2+0.9 & $6.9 \pm 2.1$ & $\sim 475$ 
	&  35.55 & $\le 25 $ & 1 \\
  B1757$-$24 & G5.4$-$1.2 & $4.6 \pm 1.4$ & $< 550$
	& 36.41 & $1.5 \pm 0.4$ & 2,3 \\
  B1853+01 &       W 44 & $3.3 \pm 1.0$ & $\sim 375$ 
	&  35.63 & $\le 1$ & 4\\
  B1951+32 &     CTB 80 & $2.4 \pm 0.6$ & $280 \pm 80$ 
	&  36.57 & $3.0 \pm 0.8$ & 5,6,7\\
\\
  B0906-49 & \nodata & $6.6 \pm 2.0$ & $\sim 60$ 
        &  35.69 & $\le 1$ & 8 \\
  B1823$-$13 & \nodata & $4.1 \pm 1.0$ &  \nodata 
	& 36.45  & $\le 40$ & 9 \\
  B1929+10   & \nodata & $0.33 \pm 0.01$ & $163 \pm 5$
	& 33.59 & \nodata & 10,11,12 \\
\\
J0537$-$6910 & N157B & $\sim 50$ (LMC) & $\sim 600$ 
	&  38.68 & $1.2 \pm 0.4$ & 13 \\
J0617.0+2221 & IC 443 & $1.5 \pm 0.6$ & $\sim 250$ 
	& $36 \pm 0.3$ & $8.5 \pm 2.0$ & 14,\dag \\
 The Mouse & G359.2$-$0.5? & $5.0 \pm 1.0$ & $\sim 180$
	& $36.3 \pm 0.3$ & $15 \pm 4$ & 15,16,\dag\\

\enddata 

\tablerefs{(1) \citet{GFGV01}; (2) \citet{FK91}; 
(3) $\mu<25$~mas~yr$^{-1}$, \citet{GF00}; (4) \citet{FGGD96}; 
(5) \citet{S87}; (6) \citet{MGB+02}; (7) \citet{HK88}; 
(8) \citet{GSFJ98};
(9) \citet{FSP96}; (10) \citet{YHH94}; (11) \citet{WLB93}; 
(12) \citet{B01}; (13) \citet{WGCD01}; (14) \citet{OCW+01}; 
(15) \citet{YB87}; (16) \citet{PK95}; 
(\dag: Not detected as a radio or X-ray pulsar.)}

\end{deluxetable}

None of the listed pulsars has a parallax measurement, forcing us to
rely on DM distances or estimates of the distance to the associated
supernova remnant for $D$. The velocities are estimated from a recent
proper motion measurement for B1951+32 \citep{MGB+02} and from the NS
age and displacement from the supernova remnant center for B1853+01
and B1643$-$43.  {\em Ad hoc} errors $\sim 25\%$ are assigned to the
latter estimates, but it is worth emphasizing that they may be larger.
For example, for B1757$-$24, a proper motion of 63--80~mas~yr$^{-1}$
has been estimated from its spindown age and displacement from the
alleged center of the remnant G5.4$-$1.2, but \citet{GF00} find a $5
\sigma$ upper limit $< 25$~mas~yr$^{-1}$ on its proper motion from
interferometric observations.  Only future interferometric measurements
of the proper motions and parallaxes of these pulsars will reduce the
large uncertainty associated with these estimates.

As before, the estimated stand-off angles (or limits) are plotted
against a combination of observable parameters, and lines of constant
density and inclination angle to the LOS are superimposed on the
plot. It is immediately apparent that the large errors on the
measurable parameters limit the useful constraints on the ISM density
that can be extracted from the plot, and to illustrate the associated
uncertainties, the location of B1757$-$24 using the erroneous proper
motion estimate of 63--80~mas~yr$^{-1}$ is also shown.  However, it is
reassuring that in spite of the large uncertainties, all the objects
are consistent with physically plausible values of the ISM density.

\begin{figure}[htf]
\plotone{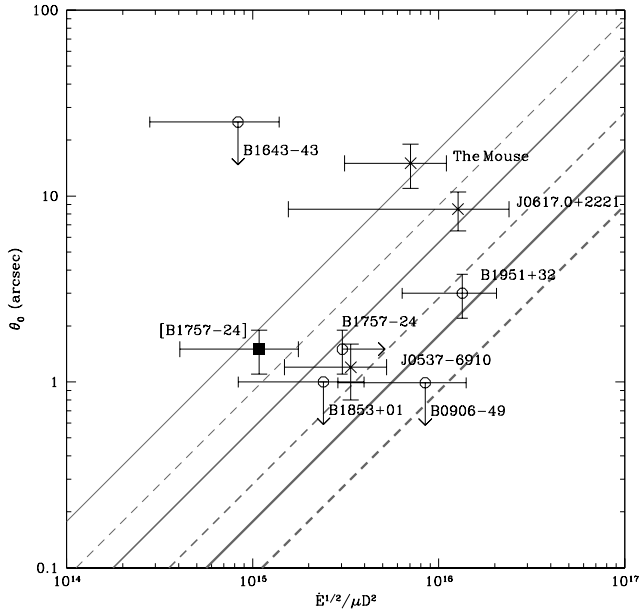}
\caption{Scaling of stand-off angle with other observable parameters
for X-ray and radio-detected bow shock nebulae,
as listed in \Tref{scale-radio}.  Circles mark radio pulsars and
crosses mark the X-ray pulsar J0537$-$6910, the radio-quiet NS
CXOU~J0617.0+2221, and the Mouse. The solid square shows B1757$-$24
when its proper motion is (erroneously) estimated from its spindown
age and displacement from the alleged supernova remnant center,
emphasizing the possible errors on this plot.  Lines of constant
number density $\nA =$ 0.01, 0.1, 1.0~cm$^{-3}$ are overplotted, for
inclination to the line of sight $i = 90\arcdeg$ (thinnest to thickest
solid lines) and $i = 45\arcdeg$ (thinnest to thickest dashed lines).}
\label{Fig:scale-radio}
\end{figure}

An unusual radio nebula has been observed for B0906$-$49
\citep{GSFJ98}, which appears to be a pulsar wind nebula generated by
a slow-moving ($\sim 60$ km \persec) pulsar and confined by the ISM
with a stand-off angle $\le 1\arcsec$. Using an analysis similar to
this work, \citet{GSFJ98} conclude that a dense ISM is required
($\nA > 2$~cm$^{-3}$).

Two other pulsars may have associated X-ray bow shock nebulae:
B1823$-$13 and B1929+10. Both of these nebulae are ROSAT
detections. For B1823$-$13, \citet{FSP96} suggest that the observed
extended component may be either the associated supernova remnant for
this young pulsar (spindown age $\sim 21$~kyr) or a bow shock nebula.
If it is a bow shock nebula, the large scale size observed for the
nebula ($\sim 40\arcsec$) implies either a low density medium or a low
NS velocity. In the absence of a proper motion estimate, the situation
cannot be resolved.  For B1929+10, ROSAT data show a diffuse X-ray
trail aligned with the pulsar proper motion, which has been explained
as synchrotron radiation from a bow shock \citep{YHH94,WLB93}. While
an improved interferometric determination of the proper motion and
parallax is available for this pulsar \citep{B01}, the stand-off angle
cannot be determined from the ROSAT data. Future observations will
complete the picture for these two pulsars.

In addition to synchrotron nebulae associated with known radio
pulsars, there are two nebulae which are plausibly bow shocks
associated with radio-quiet NS, the Mouse and the cometary nebula in
IC~443, and a bow shock nebula associated with the X-ray pulsar 
J0537$-$6910, in the LMC.

G359.23$-$0.82 (the ``Mouse'') was discovered in VLA observations by
\citet{YB87}. It appears to be a cometary wake, possibly emerging from
the circular SNR G359.2$-$0.5. Though no associated radio pulsar was
known, \citet{PK95} found an X-ray counterpart, and suggested that the
object was most likely a young NS bow shock, analogous to the one
produced by B1951+32.  They also estimate an \Edot\ $\sim 10^{36.3}$
erg \persec\ from the ROSAT X-ray spectrum.  \citet{YB87} estimate a
velocity of $\sim 300$~km~\persec\ for $D \approx 8.5$~kpc, which we
scale down to 180~km~\persec\ for $D \approx 5.0 \pm 1.0$~kpc
\citep{PK95}.  For a stand-off angle $\sim 15\arcsec$, this implies an
upper limit on the ISM density, $\nA \lesssim 0.01$~cm$^{-3}$, which
is fairly low. However, the previous caveats apply here, especially to
the estimates of velocity and \Edot.  A faint radio pulsar counterpart
has recently been discovered (F.\ Camilio, private communication),
opening up the possibility that these estimates may be improved in
future.

A similar example is provided by the cometary nebula discovered by
\citet{OCW+01} in the supernova remnant IC~443. The NS producing this
bow shock was detected by Chandra (CXOU~J0617.0+2221), along with an
X-ray counterpart to the radio synchrotron nebula.  Based on the X-ray
luminosity, $\dot{E} \sim 10^{36.3}$ erg \persec. The
distance to the associated SNR (and hence the NS) is $\sim 1.5$~kpc.
From the bow shock morphology and the ratio of the NS velocity to the
shock velocity in the SNR, \citet{OCW+01} estimate a velocity of $250
\pm 50$~km~\persec, independent of the distance.

J0537$-$6910 is a young Crab-like X-ray pulsar (16 ms period) in the
LMC, associated with the SNR N157B, which was serendipitously
discovered in RXTE observations \citep{MGZ+98}, but is not detected as
a radio pulsar.  Chandra observations show large scale diffuse
emission, which is again consistent with a ram-pressure confined
pulsar wind nebula \citep{WGCD01}.  The pulsar has a measured spindown
$\dot{E} = 4.8 \times 10^{38}$ erg \persec, and a velocity $\sim 600$
km \persec\ estimated from its age and displacement from the SNR
center.  The estimated stand-off angle $= 1\farcs2 \pm 0\farcs4$ is
consistent with the bow shock description.

The parameters for these objects are included in \Tref{scale-radio},
and they are plotted in \Fref{scale-radio}.  The uncertainties are
large, given the low confidence estimates of $D$, \Edot\ and
velocity. However, constraints on the density of the ISM can still be
obtained, and their location on the plot suggests that the bow shock
model is a suitable one for these objects.

\section{DISCUSSION AND FUTURE WORK}\label{Sec:end}

Bow shock nebulae, produced by NS moving supersonically through the
ISM, constitute unique probes of the energetic environment where NS
relativistic winds interact with the surrounding ISM. In this work, we
report multiwavelength observations of the Guitar Nebula, a
spectacular \Halpha\ bow shock nebula produced by a high velocity,
modest \Edot\ radio pulsar.  High resolution HST observations show the
detailed structure of the bright head of the nebula, confirming its
limb-brightened nature and demonstrating the existence of a
significant asymmetric component. The complex shape is best explained
by the existence of a density gradient in the medium which is not
aligned with the NS velocity vector.

A simple momentum conserving bow shock model is used to fit the HST
observations and extract the stand-off angle $\theta_0$, as well as
placing joint constraints on the inclination to the LOS, the density
of the ambient medium, and the velocity and distance of the pulsar. 
While the low S/N ratio of the current HST image precludes a more
detailed analysis, future HST data will provide more definitive
constraints for some of these parameters.

The stand-off angle $\theta_0$ and the \Halpha\ flux $F_\alpha$ scale
with other measurable parameters (\Edot, $D$, $\mu$) and unknowns
($\nH$, $X$). These scaling relationships are derived from the basic
physics of the bow shock formation.  Using the observations of the
Guitar Nebula (described here) along with published parameters for the
other known \Halpha\ bow shocks, we confirm the scaling of $\theta_0$
with the other parameters for a diverse collection of nebulae, from
simple canonical bow shocks to contorted shapes.  The scaling
relationship is used to derive upper limits on the density of the
ambient medium, independent of the inclination of the bow shock to the
LOS (which is hard to measure in a model-independent way).
Additionally, the scaling of \Halpha\ flux is demonstrated with
different combinations of measurable parameters: while the
relationship is not as constraining in this case, the observations
conform to the expected scaling within the inherent uncertainties in
the neutral fraction and extinction and the unknown density of the
ambient medium.

We have also consolidated the available parameters for radio and X-ray
bow shocks described in the literature, both for standard radio
pulsars and radio-quiet or undetected NS. The stand-off angle appears
to scale as expected for these objects as well.  However, as described
by \citet{GF00} and illustrated by \Fref{scale-radio}, the information
available for these NS is inadequate, and has large margins of
uncertainty.  VLBI measurements of the parallaxes and proper motions
of these objects will be essential for a clearer understanding of the
physics of plerionic bow shock nebulae and their correlation with
young NS and well-defined supernova remnants.

The observed anti-correlation of \Halpha\ and plerionic radio emission
from bow shocks is suggestive, though it needs to be confirmed in a
larger sample of objects.  Some young NS may be inside superbubbles
created by previous supernovae, where the neutral fraction is
insignificant.  Pre-ionization of the medium, either by the young NS,
or by the supernova in which it was born, may also suppress \Halpha\
emission by reducing the enutral fraction, leading to the prediction
that young NS with associated supernova remnants should not produce
\Halpha\ bowshocks.  Conversely, if radiation from a young NS
pre-ionizes the medium, \Halpha\ bow shocks should only be associated
with older NS.  This would be consistent with the apparent drop off in
the efficiency with which \Edot\ is converted to radio emission in
pulsar wind nebulae as a function of age \citep{GSF+00}.

Thus there are multiple avenues of future enquiry that promise to be
fruitful.  As the recently reported discovery of a bow shock nebula
for B0740$-$28 demonstrates \citep{JSG02}, deep observations of
suitably selected pulsars will probably result in the detection of
more \Halpha\ bow shocks, leading to constraints on the NS and ISM
properties, as described in this work.  Future radio and X-ray
observations will also produce more such objects, and clarify the
relationship between the NS age and environment and the type of bow
shock emission.  Additionally, further interferometric proper motions
(and parallaxes, where possible) are required to produce firmer
constraints on the properties of the ISM and the NS using known bow
shock nebulae, while theoretical models of bow shock evolution can
elucidate the differences in the behavior of bow shocks downstream
from the stand-off region.  It is also possible that old high-velocity
neutron stars, including some born in the Galactic halo, may be
identified through the serendipitous discovery of radio, X-ray or
\Halpha\ bow shock nebulae.

The Guitar Nebula, specifically, will continue to be of interest, both
from a theoretical modeling perspective, and for observations of its
evolution over time.  Monitoring of the pulsar dispersion measure and
changes in the shape of the head of the Guitar in multi-epoch high
resolution observations will provide a unique perspective on the
small-scale inhomogeneities in the interstellar medium.

\acknowledgements

We thank Joseph Lazio, Jeremy Darling and the anonymous referee for
their helpful comments.  This work is based in part on observations
made with the NASA/ESA Hubble Space Telescope, obtained at the Space
Telescope Science Institute, which is operated by the Association of
Universities for Research in Astronomy, Inc., under NASA contract NAS
5-26555. These observations are associated with proposal 5387.  The
National Radio Astronomy Observatory is a facility of the National
Science Foundation (NSF) operated under cooperative agreement by
Associated Universities, Inc.  This work at Cornell was supported in
part by NSF grant AST 9819931, and made use of NASA's Astrophysics
Data System Abstract Service and the {\tt arXiv.org} astro-ph preprint
service.

\end{document}